\documentclass[aps,prl,reprint,noshowpacs,superscriptaddress]{revtex4-2}  

\usepackage[utf8]{inputenc}
\usepackage{natbib}
\usepackage[pdftex,dvipsnames]{xcolor}
\usepackage{graphicx}
\usepackage{amsmath}
\usepackage{svg}
\usepackage{amssymb}
\usepackage{comment}
\usepackage{bm}        
\usepackage{txfonts}
\usepackage{multirow}
\usepackage{siunitx}
\usepackage{xargs}                      
\usepackage{microtype}

\usepackage[normalem]{ulem} 

\usepackage[%
  colorlinks=true,
  urlcolor=blue,
  linkcolor=blue,
  citecolor=blue
]{hyperref}

\DeclareSIUnit\bar{bar}

\begin{document}

\title{Fast quantum squeezing of a nanomechanical oscillator with an inverted potential} 


\author{Oscar Schmitt Kremer}
\affiliation{Photonics Laboratory, ETH Z\"urich, 8093 Z\"urich, Switzerland}
\affiliation{Quantum Center, ETH Z\"urich, 8093 Z\"urich, Switzerland}
\author{Lorenzo Dania}
\affiliation{Photonics Laboratory, ETH Z\"urich, 8093 Z\"urich, Switzerland}
\affiliation{Quantum Center, ETH Z\"urich, 8093 Z\"urich, Switzerland}
\author{Lukas Novotny}
\affiliation{Photonics Laboratory, ETH Z\"urich, 8093 Z\"urich, Switzerland}
\affiliation{Quantum Center, ETH Z\"urich, 8093 Z\"urich, Switzerland}
\author{Martin Frimmer}
\affiliation{Photonics Laboratory, ETH Z\"urich, 8093 Z\"urich, Switzerland}
\affiliation{Quantum Center, ETH Z\"urich, 8093 Z\"urich, Switzerland}

\date{\today}

\begin{abstract}

Nonclassical states of nano- and micro-mechanical motion enable measurements beyond the standard quantum limit and constitute a key resource for quantum sensing and metrology.
The most strongly squeezed mechanical states to date have been generated with electromechanical platforms under cryogenic refrigeration and using reservoir engineering. 
To push mechanical systems deeper into the quantum-squeezed regime requires protocols that increase the rate at which squeezing is generated to more strongly overcome the decoherence rate at which state purity is lost. 
Here, we squeeze the 800~kHz libration mode of a silica nanoparticle optically levitated in vacuum at room temperature.
We expose our mechanical oscillator to an optically generated inverted potential, where the squeezing operation proceeds at an exponentially accelerated rate. 
We reach a squeezed quadrature variance 11~dB below the vacuum fluctuations within 250~ns.
Our protocol establishes a new paradigm for generating quantum squeezing of mechanical motion and offers a platform for quantum-enhanced sensing with massive oscillators.

\end{abstract}
\maketitle

\section{Introduction}
Quantum mechanics imposes an inevitable amount of uncertainty upon the knowledge an observer can obtain about the state of a physical system. This uncertainty limits performance in applications such as information processing or sensing~\cite{Caves1982_quantumlimits,Bergeal2010-analog}. 
Squeezed states are key resources for overcoming these limits~\cite{Degen2017_quantumSensing}. They reduce the uncertainty of one variable (which is relevant for the sensing task) at the expense of that of its conjugate counterpart, while maintaining the product of the two in accordance with the Heisenberg relation~\cite{Pezze2018_RMPquantumMetro}. 
Quantum squeezing is of particular interest when applied to nano- and micro-mechanical oscillators. 
These systems efficiently couple to various degrees of freedom, making them attractive building blocks for transduction and signal processing schemes~\cite{Chu2020_APL_perspective, Mirhosseini2020,Andrews2014_bidirectional} and as sensors for fundamental tests of physics~\cite{Chou2023_quantumHighEnergy,Carney_2021_mechQuantSens,Manley2021_PRL_SearchingVector}. 

Squeezing massive systems poses a formidable challenge due to the high thermal population of the relevant bath modes at typical mechanical frequencies. 
Nevertheless, quantum squeezing of mechanical motion has been demonstrated in a variety of cryogenic electromechanical systems, including microfabricated membranes \cite{wollman2015quantum,pirkkalainen2015squeezing,lei2016quantum} and bulk acoustic resonators coupled to superconducting qubits \cite{youssefi2023squeezed,marti2024quantum}. 
Record-high levels of squeezing by $8~$dB below the vacuum fluctuations have been demonstrated using an electromechanical drum and reservoir engineering~\cite{delaney2019measurement}. 

Next to mechanically tethered mechanical oscillators, optically levitated nanoparticles have emerged as a distinct route toward quantum control of mechanical motion~\cite{delic2020cooling,Tebbenjohanns2021,Kamba2022_optColdDamp}. A levitated particle's center-of-mass motion has been squeezed by up to 7~dB by dynamically modifying the trapping potential~\cite{rossi2025quantum}. 
Besides residual decoherence, it is the low rates at which squeezing occurs in weakly confining potentials~\cite{rossi2025quantum} or under free evolution~\cite{kamba2025quantum} that hinders preparation of more strongly squeezed states in levitated systems.

A tantalizing route to beat decoherence and reach larger squeezing is to let the oscillator evolve in an inverted potential, where squeezing proceeds at an exponential rate~\cite{RomeroIsart2017_inflationNJP}. 
Optically generated inverted potentials have been deployed to manipulate levitated oscillators in thermal states~\cite{duchan_nanomechanical_2025}.
First approaches with electrically generated inverted potentials have been constrained to the classical realm due to poor state initialization, instability in the alignment of the optical and electrical potential, and small stiffness of the inverted potential~\cite{tomassi2026accelerated,seta2026shot}. 

Here we squeeze the motion of a nanomechanical oscillator by 11~dB below the vacuum level. Our oscillator is the libration mode of an optically levitated nanoparticle initialized close to its quantum ground state by cavity cooling. 
We effect the squeezing operation by evolution in an inverted potential that is optically generated and provides rapid squeezing within 250~ns, less than one fifth of one oscillation period. 
We observe the decay of the squeezed state during its lifetime of 30 oscillation periods, limited by coupling to the optical cavity used for initialization. 
Our results bear relevance for tests of quantum physics at macroscopic scales and for quantum sensing protocols.

\section{Experimental platform}
\begin{figure}[!htb]
\includegraphics{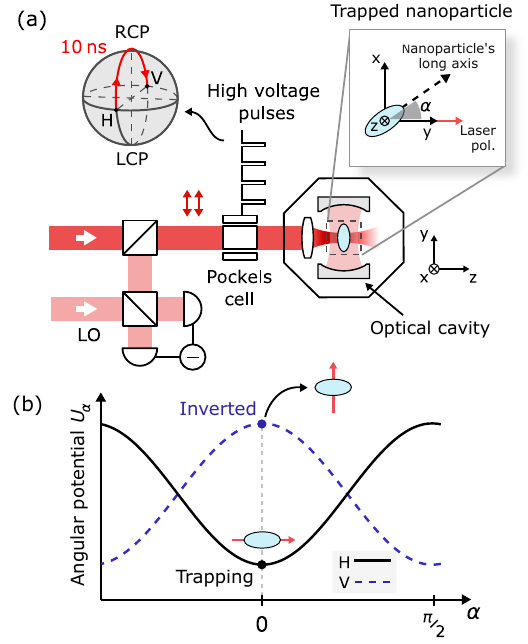}
\caption{\textbf{Optomechanical platform.} 
(a)~Experimental setup: an anisotropic nanoparticle is trapped by a horizontally ($y$) polarized tweezer and cooled to its librational ground state via coherent scattering in an optical cavity. Both the particle and the high-finesse cavity mirrors are placed inside a chamber maintained at ultrahigh vacuum and room temperature. A Pockels cell allows fast switching of the tweezer polarization from horizontal to vertical ($x$). Angular motion is monitored via homodyne measurements with a backward balanced detector. Inset figure shows an illustration of an anisotropic nanoparticle within the focal plane of the optical tweezer. The optical beam propagates along the $z$ axis and the particle long axis forms an angle $\alpha$ with respect to the tweezer polarization. 
(b)~Angular trapping potential $U_\alpha$ for a linearly polarized tweezer. For $\alpha\sim 0$, the potential is harmonic for horizontal polarization, and the particle librates around the $y$ direction. A fast switch in the tweezer polarization from horizontal to vertical places the librator in an inverted harmonic potential. H, horizontally polarized; V, vertically polarized; RCP, right-hand circularly polarized; LCP, left-hand circularly polarized; LO, local oscillator.}
\label{fig:fig1}
\end{figure}
Our experimental setup is shown in Fig.~\ref{fig:fig1}a with more details provided in Ref.~\cite{supplement}.
We trap a single anisotropic silica nanoparticle---a cluster formed by nanospheres with nominal diameter $120\,\rm{nm}$---with an optical tweezer (power $0.9\,\rm{W}$, numerical aperture $\rm{NA}=0.75$, wavelength $\lambda=1550\,\rm{nm}$). The optical trap is located in a vacuum chamber kept at a pressure of \SI{1e-9}{\milli\bar}, and at room temperature. The tweezer beam propagates along the $z$ axis and is horizontally polarized along the $y$ axis. The optical gradient force results in the oscillation frequencies $(\Omega_x, \Omega_y, \Omega_z)/(2\pi) = (185, 214, 66)\,\rm{kHz}$ for the center-of-mass motion of the nanoparticle along $x$, $y$, and $z$, respectively. 
In addition to its translational motion, an anisotropic particle also has three orientational degrees of freedom. 
In a linearly polarized tweezer, an anisotropic particle aligns its long axis to the polarization direction. Deviations from this alignment lead to a restoring torque linear in deviation angle~\cite{Hoang2016_torsional}. The associated torsional oscillations of the orientation are termed ``librations''. 
For a particle with three distinct moments of inertia, we observe three non-degenerate librational modes~\cite{kamba2023}, denoted by the angles $\gamma$, $\beta$, and $\alpha$, with librational frequencies $(\Omega_\gamma, \Omega_\beta, \Omega_\alpha)/(2\pi) = (175, 370, 810)\,\rm{kHz}$, respectively.  
In this work, we focus on the $\alpha$ mode, illustrated in Fig.~\ref{fig:fig1}a. This mode corresponds to angular oscillations of the particle's long axis in the tweezer focal ($xy$) plane \cite{rudolph2021theory}.
We detect the particle orientation by interfering the backscattered light with a local oscillator in a homodyne detection scheme.

We initialize the $\alpha$ mode close to its quantum mechanical ground state by coupling it to a high-finesse optical cavity via coherent scattering~\cite{delic2020cooling, piotrowski2023simultaneous, dania2025high, troyer2026quantum}. The cavity has linewidth $\kappa/(2\pi) = 330\,\rm{kHz}$ and resonance frequency $\Omega_c$. 
%
Throughout the duration of the experiment, the cavity resonance frequency is set to $\Omega_c   = \Omega_{\rm{tw}}+\Omega_\alpha$, where $\Omega_\text{tw}$ is the tweezer frequency.  
In this configuration, the $\alpha$ libration mode is effectively cooled by enhancing its anti-Stokes scattering resonantly by the cavity mode.
The optomechanical coupling $G$ is optimized by placing the particle at an antinode of the intracavity field \cite{dania2025high}. We reach a steady-state libration occupation value of $n_s = 0.12(6)$. The dominant source of decoherence is radiation-torque shot noise arising from the tweezer back action~\cite{Seberson2020_PRA_distribution,vanDerLaan2021_subkelvin}, which results in a phonon heating rate $\Gamma_\alpha/(2\pi) = \SI{0.5(2)}{\kilo\hertz}$. 
The homodyne detector is calibrated with sideband thermometry (for which we temporarily shift the local oscillator frequency to acquire a heterodyne spectrum), such that we can express the angle $\alpha$ in units of zero-point fluctuations (zpf) of the librator~\cite{tebbenjohanns2020motional}. 


\section{Inverted potential control}

We squeeze the $\alpha$ mode by exposing it to an inverted potential, which we generate optically. 
Recall that, in a linearly polarized optical tweezer, the $\alpha$ libration mode experiences a potential $U_\alpha \propto \sin^2(\alpha)$ \cite{rudolph2021theory}, as shown in Fig.~\ref{fig:fig1}b. Around $\alpha\approx0$ the potential is harmonic with frequency $\Omega_\alpha$. In contrast, near $\alpha\approx\pi/2$, the potential is anti-confining with frequency $\Omega_{\rm inv}=\Omega_\alpha$. 
Accordingly, for a particle initialized at $\alpha\approx0$, a rapid $\pi/2$ rotation of the tweezer polarization transfers the particle to the apex of an inverted optical potential. The resulting dynamics exponentially amplify fluctuations along one phase-space quadrature while compressing the orthogonal quadrature, thereby generating a squeezed state~\cite{RomeroIsart2017_inflationNJP}.


The polarization switch is implemented with a Pockels cell placed in the tweezer path. The cell is driven by high-voltage pulses with a rise time of approximately $\SI{10}{\nano\second}$ (much shorter than one oscillation period  $2\pi/\Omega_\alpha$). 
The pulse amplitude is optimized to achieve a polarization rotation from horizontal to vertical along a meridian line of the Poincare sphere, as illustrated by the red path in Fig.~\ref{fig:fig1}a.
The pulse duration $\tau$ controls the time spent in the anti-confining regime. 

\section{Preparation and verification of angular squeezed states}
\begin{figure*}[!htb]
\includegraphics[width=2\columnwidth]{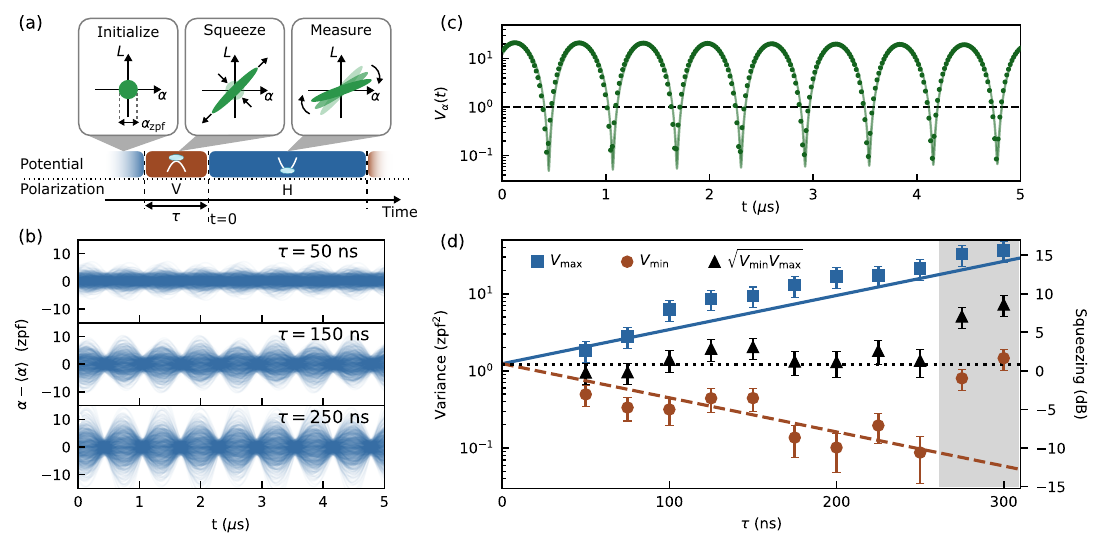}
\caption{\textbf{Preparation and verification of angular squeezed states.} 
(a)~Protocol for generating and reconstructing an angular squeezed state. The particle is initialized in the libration ground state, exposed to the inverted potential for a duration $\tau$, and recaptured in the harmonic potential. The subsequent phase-space rotation is measured by homodyne detection. The protocol is repeated 1000 times to reconstruct the state. 
(b)~Ensembles of angular time traces $\alpha(t)$ expressed in units of zero-point fluctuations (zpf) recorded after recapture for inverted-potential durations of $\SI{50}{\nano\second}$, $\SI{150}{\nano\second}$, and $\SI{250}{\nano\second}$. Each plot is obtained by band-pass filtering the raw homodyne traces around $\Omega_\alpha/(2\pi)=\SI{810}{\kilo\hertz}$ and subtracting the mean ensemble average $\langle\alpha\rangle$. 
(c)~Time evolution of the inferred angular variance $V_\alpha(t)$ after a pulse duration $\tau=\SI{250}{\nano\second}$. The dashed line indicates the zero-point variance. The oscillation at $2\Omega_\alpha$ results from the rotation of the squeezed state in phase space. The solid line is a fit to the harmonic evolution, from which we extract the covariance matrix at recapture. 
(d)~Anti-squeezed ($V_{\rm max}$, blue circles) and squeezed ($V_{\rm min}$, red squares) quadrature variances at the moment of recapture and as a function of $\tau$. The uncertainty product $\sqrt{V_{\rm min}V_{\rm max}}$ (black triangles) remains approximately constant up to $\tau=\SI{250}{\nano\second}$. At this duration, we measure $\SI{11(3)}{\decibel}$ of squeezing below the zero-point level and $\SI{13(1)}{\decibel}$ of anti-squeezing. 
For longer $\tau$ (gray shaded region), the measurements deviate from the expectation due to technical noise.
Colored lines (solid for $V_\text{max}$, dashed for $V_\text{min}$) are theory predictions based on the independently measured initial state variance (given by $2n_s+1$) and $\Omega_{\rm inv}$. The black dotted line corresponds to the steady state variance under cavity cooling, less than \SI{1}{\decibel} above the zero point variance.}
\label{fig:fig2}
\end{figure*}

We generate angular squeezed states with the experimental protocol illustrated in Fig.~\ref{fig:fig2}a.
With the tweezer horizontally polarized (along $y$), we initialize the $\alpha$ libration mode to its quantum ground state by cavity cooling. We expose the particle to the inverted potential for a duration $\tau$ by switching the tweezer polarization to vertical (along $x$). We denote the time $t=0$ as the moment when the polarization direction is switched back to horizontal. 
Therefore, at $t=0$, the particle is recaptured in the harmonic potential, such that the squeezed state generated in the inverted potential rotates in phase space at frequency $\Omega_\alpha$. 
This phase space is spanned by the angular orientation coordinate $\alpha$ and the angular momentum $L$ of the librator.
We characterize the librator's state by the homodyne measurement of $\alpha(t)$ collected at time $t>0$. The protocol is repeated 1000 times for each value of $\tau$ to build the phase-space distribution of the state. The repetition rate of the experiment is set to 1~Hz to ensure that all degrees of freedom are cooled back to the same initial state before the next realization.

Figure~\ref{fig:fig2}b shows ensembles of measured time traces of the particle's orientation angle $\alpha$ during the first \SI{5}{\micro\second} after recapture for $\tau=50$~ns (top), $150$~ns (middle), and $\SI{250}{\nano\second}$ (bottom). Each trace is obtained by band-pass filtering the calibrated homodyne signal around $\Omega_\alpha/2\pi$ with a bandwidth of \SI{15}{\kilo\hertz}, chosen to reject other motional modes. For clarity, we subtract the ensemble mean $\langle \alpha\rangle$. 
For all values of $\tau$, we observe that the spread of the individual timetraces (i.e., the variance of $\alpha$) oscillates in time at a frequency $2\Omega_\alpha$. This observation is consistent with a squeezed phase-space distribution rotating at frequency $\Omega_\alpha$ in the $\alpha$-$L$ plane.
Furthermore, we observe that the amplitude of the oscillations of the variance increases with the time spent in the inverted potential $\tau$. This observation is in line with the expectation that the level of squeezing grows with $\tau$.

To quantitatively analyze our data, we investigate the variance $\tilde V_\alpha(t)$ of the measured time traces.
This measured variance has two contributions. The first contribution is the transduced motion of the particle $V_\alpha$. The second contribution is the unavoidable measurement imprecision $V_n$ from photon shot noise at the optical detector. We determine the measurement imprecision $V_n=1.34(1)$ in an independent measurement by integrating the imprecision noise in the measurement bandwidth.  
We then extract the variance of the particle's angular motion by computing $V_\alpha=\tilde{V}_\alpha-V_n$~\cite{youssefi2023squeezed,rossi2025quantum}.
Figure~\ref{fig:fig2}c shows $V_\alpha(t)$ for $\tau=250\,\rm{ns}$. 
We observe harmonic oscillations of $V_\alpha(t)$ at the frequency $2\Omega_\alpha$. 
Importantly, the measured variance periodically falls below the zero-point level (dashed line), demonstrating quantum squeezing of the angular motion.

To further test our understanding, we model $V_\alpha(t)$ for a squeezed harmonic oscillator under cavity cooling, which yields
\cite{supplement}
\begin{multline}\label{eq:variance_evolution}
    V_\alpha(t) = \text{e}^{-\gamma_c t}[V_{\alpha,0}\cos^2(\Omega_\alpha t) + V_{L,0}\sin^2(\Omega_\alpha t) \\ +  C_0\sin(2\Omega_\alpha t)]
    + V_{\alpha, \infty}(1-\text{e}^{-\gamma_c t}).
\end{multline}
Here, $V_{\alpha,0}$ is the angular variance, $V_{L,0}$ the angular-momentum variance, and $C_0$ their covariance, each at recapture $(t=0)$. 
Furthermore, $V_{\alpha,\infty}$ is the steady-state variance of the librator while coupled to, and therefore cooled by, the cavity at the rate $\gamma_c$. 
The value of this steady-state variance is given by the steady-state phonon occupation under cavity cooling $n_s$ according to $V_{\alpha,\infty}=2n_s+1$. The cavity-cooling rate $\gamma_c/(2\pi)=\SI{4.1(3)}{\kilo\hertz}$ is extracted from the linewidth of the motional spectrum under steady-state cooling.
Fitting the experimental data $V_\alpha(t)$ to Eq.~\eqref{eq:variance_evolution} provides as fit parameters the elements of the covariance matrix at $t=0$:
\begin{equation}
    \mathbf{V}_0  = \begin{bmatrix}
        V_{\alpha,0} & C_0 \\ 
        C_0 & V_{L,0} 
    \end{bmatrix}.
\end{equation}
Diagonalizing $\mathbf{V}_0$ yields the anti-squeezed and squeezed quadrature variances, $V_{\rm max}$ and $V_{\rm min}$, respectively.

Figure~\ref{fig:fig2}d shows $V_{\rm max}$ and $V_{\rm min}$ as a function of evolution time $\tau$ in the inverted potential. 
We observe that $V_{\rm max}$ grows and $V_{\rm min}$ shrinks exponentially. Only for long evolution times in the inverted potential ($\tau>250$~ns) does $V_{\rm min}$ depart from the exponential scaling.
We add as lines the expected evolution of the variances according to our model (no free fit parameter) of a harmonic oscillator in an inverted potential $V_{\rm max(min)}(\tau)=V_{\alpha,\infty}\text{e}^{\pm\Omega_{\rm inv} \tau}$. The model prediction matches the data such that, indeed, our librator's state is exponentially squeezed at a rate given by the stiffness of the inverted potential $\Omega_\alpha$. 
At $\tau=\SI{250}{\nano\second}$, we measure the strongest squeezing, which is $\SI{11(3)}{\decibel}$ below the zero-point fluctuations, accompanied by $\SI{13(1)}{\decibel}$  of anti-squeezing.

To investigate the quality of our squeezing operation further, we extract from our measurements the uncertainty product $\sqrt{V_{\rm min}V_{\rm max}}$. Its inverse, the state purity $\mathcal{P}=(V_{\rm min}V_{\rm max})^{-\frac{1}{2}}$, is bounded by $\mathcal{P}\le1$ as a consequence of the Heisenberg uncertainty relation. 
Figure~\ref{fig:fig2}d shows the measured values of $\sqrt{V_{\rm min}V_{\rm max}}$ as black triangles. 
Up to $\tau=\SI{250}{\nano\second}$, the uncertainty product remains, within the errorbars, at the value established by cavity cooling (dashed black line). 
In particular, at $\tau=\SI{250}{\nano\second}$, corresponding to the maximum observed squeezing, we infer the state purity $\mathcal{P}=0.74$. 
This observation indicates that, for $\tau\le 250~$ns the squeezing process introduces negligible added uncertainty to the state. 

For longer inverted-potential durations, we observe a degradation in squeezing accompanied by a decrease in purity. 
To elucidate the observed limitation of our protocol, let us consider the backaction limit, where decoherence arises purely from radiation-torque shot noise in the inverted potential. 
In this backaction-limited scenario, we would expect squeezing up to $\SI{27}{\decibel}$ at $\tau=\SI{790}{\nano\second}$. 
We therefore conclude that our protocol is currently limited by a technical noise process. We suspect that the degradation in purity for $\tau>250~$ns arises from coupling to another libration mode (which is not cooled efficiently by the cavity). We conjecture that this mode heats up sufficiently during long $\tau$ to perturb the $\alpha$ mode via a non-linear mechanical coupling process. This may be remedied by initialization of all degrees of freedom to a low occupation prior to squeezing~\cite{kamba2023}.


\section{Evolution of a squeezed state under cavity cooling}

Finally, we investigate the relaxation dynamics of the squeezed state for long evolution times $t$ after retrapping in the confining potential. 
%
%
%
\begin{figure}[!tb]
\includegraphics{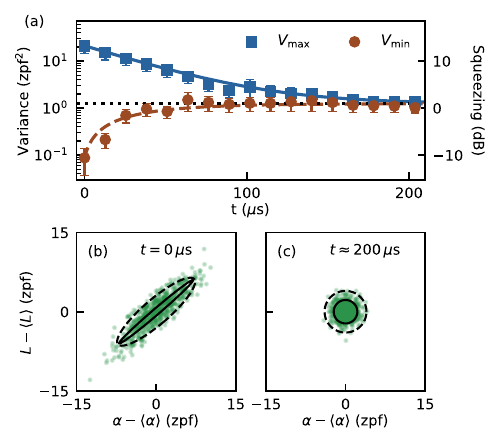}
\caption{\textbf{Evolution of a squeezed state under cavity cooling.}
(a)~Measured variances of the anti-squeezed ($V_{\rm max}$, blue circles) and squeezed ($V_{\rm min}$, red squares) quadratures as a function of time after preparation of a squeezed state with $\tau=\SI{250}{\nano\second}$. The dotted black line indicates the steady-state variance under cavity cooling. Solid and dashed lines are exponential functions of the form $[V_{\alpha,\infty}\pm V_{\rm max(min)}(0)]e^{-\gamma_ct}+V_{\alpha,\infty}$, with decay time $1/\gamma_c=\SI{38}{\micro\second}$ extracted from steady-state cavity cooling measurements (no fits). 
(b)~Reconstructed phase-space distribution immediately after squeezing. 
(c)~Reconstructed phase-space distribution after \SI{200}{\micro\second} of evolution under cavity cooling. Dashed contours in (b) and (c) correspond to the measured distributions, while solid contours show the inferred distributions after measurement shot-noise subtraction.}
\label{fig:fig3}
\end{figure}
We prepare a squeezed state using an inverted-potential duration of $\tau=\SI{250}{\nano\second}$. Upon recapture in the confining potential, we track the quadrature dynamics as a function of time $t$ spent in the harmonic trap. Figure~\ref{fig:fig3}a shows the evolution of the squeezed variance $V_{\rm min}(t)$ and anti-squeezed variance $V_{\rm max}(t)$ over \SI{200}{\micro\second}.
We observe both quadratures converging from their squeezed values to the same steady-state value. 
%
Theory predicts that both quadratures decay exponentially to $V_{\alpha,\infty}$, the steady-state ground-state variance, at the rate $\gamma_c$ according to~\cite{supplement}
\begin{equation}\label{eq:var_decay}
V_{\rm max(min)}(t)=\left[V_{\rm max(min)}(0)-V_{\alpha,\infty}\right]e^{-\gamma_ct}+V_{\alpha,\infty}.
\end{equation}
This result is expected upon recalling that, during the entire experiment, the detuning of the tweezer relative to the cavity is set to cool the libration mode. 
We plot Eq.~\eqref{eq:var_decay} in Fig.~\ref{fig:fig3}a with $V_{\alpha,\infty}$ and $\gamma_c$ extracted from steady-state measurements under cavity cooling and $V_\text{max(min)}(0)$ set to the respective measured values (no free fit parameter), showing good agreement with our measured variances.
The squeezed state persists for $\Omega_\alpha/(2\pi\gamma_c)\approx30$ libration periods.   

To further illustrate the squeezed state, we show in Figs.~\ref{fig:fig3}b and c the reconstructed phase-space distributions immediately after squeezing ($t=0$) and after $t=\SI{200}{\micro\second}$ of evolution under cavity cooling, respectively. 
Each datapoint represents the angle and angular-momentum value measured during one iteration of the experiment.
The initially squeezed distribution gradually relaxes back to the steady-state ground state under the action of the cavity-induced dissipation.

In our experiment, the lifetime of the squeezed state is limited by the optomechanical interaction with the cavity field. 
One could overcome this limit by judiciously detuning the cavity mode away from the anti-Stokes sideband of the librator in the measurement-step of the protocol. In this situation, the lifetime of the squeezed state would be given by the backaction heating rate. We estimate a recoil-limited lifetime of $\Gamma_\alpha^{-1}\approx\SI{0.3}{\milli\second}$, corresponding to $\Omega_\alpha/(2\pi\Gamma_\alpha)=240$ libration periods, roughly one order of magnitude longer than the measured cavity-limited decay time. 


\section{Conclusions}  
In conclusion, we have squeezed the mechanical motion of an optically levitated nanomechanical librator below the vacuum fluctuations by 11(3)~dB in variance.
As compared to mechanically tethered oscillators, where the strongest levels of squeezing have been achieved by reservoir engineering in cryogenic environments~\cite{wollman2015quantum,pirkkalainen2015squeezing, lei2016quantum, delaney2019measurement}, our protocol has exploited the rapid modulation of the trapping potential and operates at room temperature. 
In contrast to earlier approaches to squeezing in levitation, which deployed different confining potentials~\cite{rossi2025quantum} or free evolution~\cite{kamba2025quantum}, we have leveraged the exponentially accelerated squeezing provided by an inverted potential~\cite{RomeroIsart2017_inflationNJP}. 
Importantly, we have generated the inverted potential optically, which offers two significant advantages compared to approaches using static electric fields~\cite{tomassi2026accelerated,seta2026shot}. 
First, the large stiffness of the optical inverted potential leads to fast squeezing, less than one fifth of the period of our oscillator. Second, the all-optical approach is insensitive to electric stray fields and displacement noise resulting from misalignment of the optical and electrical potential~\cite{seta2026shot}. 



Our results bear relevance for different research directions. 
Access to a mechanical squeezed state unlocks the potential of quantum-enhanced sensing protocols pioneered with the center-of-mass motion of trapped ions~\cite{Burd2019_science} and recently adapted to levitated nanoparticles~\cite{skrabulis2026nanomechanical}. 
Such levitated sensors are promising candidates to identify weakly interacting elementary particles in collision experiments~\cite{moore2021searching,tseng2025search,Carney2023_searchMassiveNeutrinos}. Exploiting the angular degree of freedom of a levitated object as an additional resource is an enticing prospect. 

Besides metrology, squeezed states also play a key role for testing quantum mechanics at large scales~\cite{GonzalezBallestero2021,Bassi2013_RMP_collapse}. 
In proposals to create center-of-mass superpositions of massive objects~\cite{RomeroIsart2011_PRA_quantumSuperpos,romero2011large}, squeezing provides state delocalization while maintaining purity to expose the wavefunction to non-linear forces for the generation of non-Gaussian states. 
Our work is a step towards realizing analogous protocols with angular degrees of freedom. 
Furthermore, quantum control over angular degrees of freedom offers the potential to explore macroscopic quantum effects genuine to rotors, such as orientational quantum revivals and quantum-persistent tennis-racket flips~\cite{Stickler2021_quantumRotations,Ma2020_PRLquantumpersistent,Stickler2018_NJPprobingMacro}.

\textit{Acknowledgments.}
We thank J. A. Zielińska, J. Gao, and the trappers of the Photonics Lab for stimulating discussions. L.D. acknowledges support from the Quantum Center Research Fellowship and the Dr Alfred and Flora Spälti Fonds.
This research has been supported by the Swiss SERI Quantum Initiative (Grants No.\ UeM019-2 and No.\ UeM029-3), and the Swiss National Science Foundation (Grant No.\ 212599).

\bibliography{bibliography}

\end{document}